\documentclass[12pt]{article}
\usepackage{epsfig,amssymb,amsmath,bbm,psfrag}

\catcode`\@=11
\DeclareMathAlphabet{\mathpzc}{OT1}{pzc}{m}{it}
\newcommand{\res}{\mathop{\rm Res}}

\font\cmss=cmss12 
\def\1{\hbox{{1}\kern-.25em\hbox{l}}}
\def\bfZ{\relax{\hbox{\cmss Z\kern-.4em Z}}}
\def \be  {\begin{equation}}
\def \ee  {\end{equation}}
\def \ba  {\begin{eqnarray}}
\def \ea  {\end{eqnarray}}
\def \baa {\begin{eqnarray*}}
\def \eaa {\end{eqnarray*}}
\def \bb  {\begin {thebibliography} }
\def \eb  {\end{thebibliography}}
\def \lab #1 {\label{#1}}

\newcommand\re[1]{(\ref{#1})}

\def \matrix #1 {\left(\begin{array}{cc} #1 \end{array}\right)}

\def \tr {\mathop{\rm tr}\nolimits}
\def \res{\mathop{\rm res}\nolimits}

\newcommand{\as}{\ifmmode\alpha_{\rm s}\else{$\alpha_{\rm s}$}\fi}
\newcommand{\asbar}{\ifmmode\bar{\alpha}_{\rm s}\else{$\bar{\alpha}_{\rm
s}$}\fi}

\newcommand{\ft}[2]{{\textstyle\frac{#1}{#2}}}

\font\cmss=cmss12 
\def\inbar{\,\vrule height1.5ex width.4pt depth0pt}
\def\IC{\relax\hbox{$\inbar\kern-.3em{\rm C}$}}
\def\IZ{\relax{\hbox{\cmss Z\kern-.4em Z}}}
\def\IR{{\hbox{{\rm I}\kern-.2em\hbox{\rm R}}}}

\def\IP{{\hbox{{\rm I}\kern-.2em\hbox{\rm P}}}}
\def\II{\hbox{{1}\kern-.25em\hbox{l}}}

\newbox\lett\newdimen\lheight\newdimen\lwidth
\def\ontop#1#2{\setbox\lett=\hbox{#2}\lheight\ht\lett
\multiply\lheight by 12 \divide\lheight by 10\relax%
\lwidth\wd\lett \multiply\lwidth by 8 \divide\lwidth by 10\relax #2\kern-
\lwidth%
\raise\lheight\hbox{{$\scriptstyle #1$}}\kern.1ex}

\def\XXint#1#2#3{{\setbox0=\hbox{$#1{#2#3}{\int}$}
     \vcenter{\hbox{$#2#3$}}\kern-.5\wd0}}

\begin{document}

\begin{titlepage}

\thispagestyle{empty}

\vskip2cm

\centerline{\large \bf Baxter equation beyond wrapping}

\vspace{1cm}

\centerline{\sc A.V. Belitsky}

\vspace{10mm}

\centerline{\it Department of Physics, Arizona State University}
\centerline{\it Tempe, AZ 85287-1504, USA}

\vspace{1cm}

\centerline{\bf Abstract}

\vspace{5mm}

The Baxter-like functional equation encoding the spectrum of anomalous dimensions of Wilson
operators in maximally supersymmetric Yang-Mills theory available to date ceases to work
just before the onset of wrapping corrections. In this paper, we work out an improved
finite-difference equation by incorporating nonpolynomial effects in the transfer matrix
entering as its ingredient. This yields a self-consistent asymptotic finite-difference
equation valid at any order of perturbation theory. Its exact solutions for fixed spins and
twists at and beyond wrapping order give results coinciding with the ones obtained from the
asymptotic Bethe Ansatz. Correcting the asymptotic energy eigenvalues by the L\"uscher term,
we compute anomalous dimensions for a number of short operators beyond wrapping order.

\end{titlepage}

\setcounter{footnote} 0

\newpage

\pagestyle{plain}
\setcounter{page} 1

{\bf 1. Asymptotic Baxter equation and wrapping.} To date \cite{Utr08} there is a significant
body of data which suggests that the spectrum of all anomalous dimensions in planar maximally
supersymmetric gauge theory can be computed overcoming complicated calculations Feynman diagrams.
This finding \cite{Lip97,MinZar03,BeiSta03,BelDerKorMan04} generalizes previous observations
that the spectrum of one-loop maximal-helicity gluon $X = F_{+\perp}$ operators in pure gauge
theory,
\be
\label{WilsonOperator}
\mathcal{O} = \tr \{ D^{N_1}_+ X (0) D^{N_2}_+ X (0) \dots D^{N_L}_+ X (0) \}
\, ,
\ee
can be calculated by identifying the dilation operator with the Hamiltonian of a noncompact Heisenberg
magnet \cite{BraDerMan98,Bel98}. The correspondence works by placing the elementary fields $X(0)$
of the Wilson operator on the spin-chain sites and identifying spin generators with the ones of the
collinear $SL(2)$ subgroup of the (super)conformal group. The noncompactness of the spin chain is
a consequence of the fact that there are infinite towers of covariant derivatives $D_+$ acting on
those fields. The $sl(2)$ subsector of the maximally supersymmetric gauge theory, which we study in
this paper, is spanned by the Wilson operators \re{WilsonOperator} with the elementary complex scalar
field $X = \phi^1 + i \phi^2$ \cite{BeiSta03}.

The one-loop integrable structure was generalized to all orders in 't Hooft coupling $g^2 =
g^2_{\rm\scriptscriptstyle YM} N_c/(4 \pi^2)$ and, though one is currently lacking a putative
spin-chain picture for the dilatation operator, a set of Bethe Ansatz equations was put forward
which survives a number of nontrivial spectral checks \cite{BeiSta05}. However, these equations
allow one to calculate anomalous dimensions for Wilson operators as long as the order of the
perturbative expansion does not exceed the length of the operator. Namely, when the
interaction of the spins in the chain start to wrap around it, the aforementioned equations start
to fail \cite{AmbJanKri05,KotLipRejStaVel07}. Thus the true anomalous dimension for the Wilson
operators are given a sum of two terms
\be
\gamma (g) = \gamma^{(\rm asy)} (g) + \gamma^{(\rm wrap)} (g)
\, .
\ee
The first contribution is determined by the solution to the asymptotic Bethe Ansatz equations and
can be written as \cite{Bel06}
\be
\label{AllOrderAD}
\gamma^{(\rm asy)} (g) = i g^2 \int_{- 1}^1 \frac{dt}{\pi} \sqrt{1 - t^2}
\left(
\ln \frac{
Q \left( + \ft{i}{2} - g t \right)
}{
Q \left( - \ft{i}{2} - g t \right)
}
\right)^\prime
\, ,
\ee
in terms of a polynomial with zeroes determined by the Bethe roots $u_k$
\be
Q (u) = \prod_{k = 1}^N (u - u_k(g))
\, .
\ee
It has to be supplemented by the condition of the vanishing quasimomentum
\be
i \vartheta
=
\frac{1}{\pi}
\int_{- 1}^1 \frac{dt}{\sqrt{1 - t^2}}
\ln \frac{Q (+ \ft{i}{2} - g t)}{Q (- \ft{i}{2} - g t)}
= 0
\, ,
\ee
in order to pick out only cyclic physical states. For vanishing 't Hooft coupling, $u_k (0) =
u_{k,0}$ coincide with the Bethe roots of the short-range $sl(2)$ XXX spin chain. However, at
higher order of perturbation theory they acquire a nontrivial dependence on the 't Hooft coupling,
and so does the Baxter function $Q (u) = Q_0 (u) + g^2 Q_1 (u) + \dots$. The Baxter polynomial
$Q(u)$ is a function of the spectral parameter and it is determined as a solution to a
finite-difference equation known as the asymptotic Baxter equation
\be
\label{BaxterEquation}
(x^+)^L {\rm e}^{\sigma_+ (u^+) - \Theta (u^+)} Q (u + i)
+
(x^-)^L {\rm e}^{\sigma_- (u^-) - \Theta (u^-)} Q (u - i)
=
t (u) Q (u)
\, .
\ee
The dressing factors accompanying the polynomial depend on the renormalized spectral parameter
$x = x [u] = \ft12 (u + \sqrt{u^2 - g^2})$ with the used notation $x^\pm = x [u^\pm]$. The
exponents $\sigma$ and $\Theta$ encode nontrivial dynamics of the long-range field-theoretical
``spin chain'', with the first one
\be
\label{Sigmap}
\sigma_\pm (u) =
\int_{-1}^1
\frac{d t}{\pi}
\frac{\ln Q (\pm \ft{i}{2} - g t)}{\sqrt{1 - t^2}}
\left( 1 - \frac{\sqrt{u^2 - g^2}}{u + g t} \right) \, ,
\ee
partially responsible for the renormalization of the conformal spin at higher orders of
perturbation theory, and the second
\ba
\Theta (u)
\!\!\!&=&\!\!\!
- 8 i \sum_{r=2}^\infty \sum_{s = r+1}^\infty
\left( \frac{g}{2} \right)^{r + s - 2} C_{rs} (g)
\int_{-1}^1 \frac{dt}{\pi} \sqrt{1 - t^2}
\left( \ln \frac{Q (+ \ft{i}{2} - g t)}{Q (- \ft{i}{2} - g t)} \right)^\prime
\nonumber\\
&&\qquad\qquad\qquad\qquad\qquad\times
\left\{
\left(- \frac{2}{g} \right)^{s - 2} \frac{U_{s - 2} (t)}{x^{r - 1}}
-
\left(- \frac{2}{g} \right)^{r - 2} \frac{U_{r - 2} (t)}{x^{s - 1}}
\right\}
\, ,
\ea
providing smooth interpolation between the weak and strong-coupling expansions \cite{BeiEdeSta06}.
Here we presented the $\Theta$-phase in a form of an infinite expansion with the transcendental
coefficients
\be
C_{rs} (g) = \sin(\ft{\pi}{2} (s - r))
\int_0^\infty d v \frac{J_{r - 1} (gv) J_{s - 1} (gv)}{v ({\rm e}^v - 1)}
\, ,
\ee
which is the most suitable form for perturbative analyses we perform in this paper.

The onset of wrapping corrections starts from the order $g^{2(L + 2)}$ for the Wilson operators
\re{WilsonOperator} \cite{AmbJanKri05,KotLipRejStaVel07}. Thus the first few orders of the
corresponding anomalous dimensions are free from these complications $\gamma^{(\rm asy)} = g^2
\gamma_0 + g^4 \gamma_1 + \dots$ and can be determined efficiently from the Baxter equation
\re{BaxterEquation}, $\gamma_\ell = \gamma_\ell^{(\rm asy)}$ for $\ell \leq L + 1$. The first
wrapping correction to the anomalous dimensions is given by a multiparticle L\"uscher formula,
which was recently conjectured and tested for the Konishi operators and four-loop twist-two
($L=2$) operators in Refs.\footnote{Here we restored the multiplicative factor
$|Q_0 (\ft{i}{2})|^2$ required to make the wrapping correction invariant under rescaling of
the Baxter function by an arbitrary constant, $Q \to \lambda Q$, allowed by the homogeneous
Baxter equation. Its apparent absence in Ref.\ \cite{BajJanLuk08} is due to a particular
normalization chosen in that paper. We would like to thank Romuald Janik for correspondence
on this point.} \cite{BajJanLuk08},
\be
\label{WrappingGamma}
\gamma^{(\rm wrap)} (g)
=
- g^4 \gamma_0^2
i |Q_0 (\ft{i}{2})|^2
\sum_{n = 1}^\infty \res\limits_{z = i n}
\left(\frac{g^2}{z^2 + n^2} \right)^L \frac{T^2 (z, n)}{R (z, n)}
+
\mathcal{O} \left( g^{2 (L + 3)} \right)
\, .
\ee
It is written in terms of
\be
R (z, n) =
Q_0 \left( \ft12 z - \ft{i}{2} (n - 1) \right)
Q_0 \left( \ft12 z + \ft{i}{2} (n - 1) \right)
Q_0 \left( \ft12 z + \ft{i}{2} (n + 1) \right)
Q_0 \left( \ft12 z - \ft{i}{2} (n + 1) \right)
\, ,
\ee
and the function
\be
T (z, n) = \sum_{m = 0}^{n - 1}
\frac{
Q_0 \left( \ft12 z - \ft{i}{2} (n - 1) + i m \right)
}{
\left[ \left( m - \ft12 n \right) - \ft{i}{2} z \right]
\left[ \left( m + 1 - \ft12 n \right) - \ft{i}{2} z \right]
}
\, ,
\ee
which also contains a kinematical pole at $z = i n$ in addition to the ones displayed
explicitly in Eq.\ \re{WrappingGamma}.

{\bf 2. Transfer matrix revisited.} The original proposal \cite{Bel06} for the transfer
matrix $t (u)$, entering the right-hand side of the Baxter equation, as a polynomial of
order $L$ in the bare spectral parameter $u$ yields an inconsistent equation one order
before the wrapping corrections set in. Since the Baxter equation has a number of
advantages over the Bethe Ansatz equations,---with relative simplicity in its diagonalization,
straightforward asymptotic solution for large values of quantum numbers of the Wilson
operators, etc. being a few,---one has to seek for modifications of Eq.\ \re{BaxterEquation}
which yield a self-consistent equation at any order of perturbation theory. A systematic
inspection demonstrates that the aforementioned limitations can be easily overcome by merely
modifying analytical properties of the transfer matrix and assuming the following ansatz for it
\be
t (u)
=
\Re{\rm e} ( x^+ )^L
\left( 2 + \sum_{k \geq 1} \mathfrak{Q}_k (g) \Re{\rm e} ( x^+ )^{- k} \right)
-
\sum_{k \geq 1} \mathfrak{R}_k (g) \Im{\rm m} ( x^+ )^{- k}
\, .
\ee
Here the upper limits in the sums depend on the length of the operator in question and the
order of perturbation theory under consideration. Here, both sets of charges, $\mathfrak{Q}_k$
and $\mathfrak{R}_k$, admit an infinite-series expansion in the 't Hooft coupling constant,
\ba
\mathfrak{Q}_k
\!\!\!&=&\!\!\! \mathfrak{Q}_k^{[0]} + g^2 \mathfrak{Q}_k^{[1]} + g^4 \mathfrak{Q}_k^{[2]}
+ \dots
\, , \\
\mathfrak{R}_k
\!\!\!&=&\!\!\! \mathfrak{R}_k^{[0]} + g^2 \mathfrak{R}_k^{[1]} + g^4 \mathfrak{R}_k^{[2]}
+ \dots
\, . \nonumber
\ea
Only the leading order charges $\mathfrak{Q}_{k \leq L}^{[0]}$ are related to the integrals
of motion of the periodic short-range $sl(2)$ spin chain for $g = 0$, with $\mathfrak{Q}_2^{[0]}$
being expressed in terms of the eigenvalues of quadratic Casimir operator of the collinear conformal
algebra,
\be
\mathfrak{Q}_2^{[0]} = - (N + \ft12 L) (N + \ft12 L - 1) - \ft{1}{4} L (L - 2)
\, ,
\ee
and $\mathfrak{Q}_1^{[0]} = 0$ vanishing to accommodate the tracelessness of the fundamental
generators. On the other hand, the primary purpose of the compensatory charges $\mathfrak{R}_k$
and $\mathfrak{Q}_{k > L}$ is to eliminate non-polynomial terms arising in the left-hand
side of the finite difference equation \re{BaxterEquation} stemming from the expansion of
the renormalized rapidity parameter and dressing factors in Taylor series in the 't Hooft coupling.
These emerge only at higher orders of perturbation theory, with $\mathfrak{R}_k^{[\ell]}$ being
nonvanishing starting only from the wrapping order,
\be
\label{NonvanishR}
\mathfrak{R}_k^{[\ell < L + 1]} = 0
\, .
\ee
They can be consistently found by matching both sides of Eq.\ \re{BaxterEquation} in terms of
the Baxter functions $Q_\ell (u)$ determined at previous orders of perturbation theory.

Though it appears that the finite-difference equation \re{BaxterEquation} contains two unknowns,
i.e., the Baxter function $Q(u)$ (or Bethe roots) and the transfer matrix $t (u)$ (or the charges),
it allows to fix both. To simplify its solution, one can use the values of the charges $\mathfrak{Q}_1$
and $\mathfrak{Q}_2$ as inputs, both of which can be determined to all order of perturbation theory
\ba
\label{q1-def}
\mathfrak{Q}_{1} (g) \!\!\!&=&\!\!\! \Delta^{(1)}_+ (g) + \Delta^{(1)}_- (g)
\, , \\
\label{q2-def}
\mathfrak{Q}_2 (g)
\!\!\!&=&\!\!\!
-
\mathbb{C}_2 (g)
+
\ft14 \left( \Delta^{(1)}_+ (g) + \Delta^{(1)}_- (g) \right)^2
+
\Delta^{(2)}_+ (g) + \Delta^{(2)}_- (g)
+
\ft14 L (L - 2)
\, .
\ea
The first term in $\mathfrak{Q}_2 (g)$ being the renormalized quadratic Casimir of the $sl(2)$
algebra
\be
\mathbb{C}_2 (g)
\equiv
\left( N + \ft12 L + \ft12 \gamma^{(\rm asy)} (g) \right)
\left( N + \ft12 L + \ft12 \gamma^{(\rm asy)} (g) - 1 \right)
\, ,
\ee
with the scaling dimension $(N + \ft12 L)$ of the Wilson operators shifted by their anomalous
dimensions,
\be
\label{AsyAD}
\gamma^{(\rm asy)} (g)
= - 2 i \left( \Delta^{(1)}_+ (g) - \Delta^{(1)}_- (g) \right)
\equiv \gamma^+ (g) - \gamma^- (g)
\, .
\ee
The above charges are written in terms of the first two leading coefficients of the large-$u$
expansion of the dressing factors $\sigma_\pm$ and $\Theta$,
\be
\Delta^{(\alpha)}_\pm =
\int_{- 1}^1 \frac{d t}{\pi}
\sqrt{1 - t^2}
\left\{
w^{(\alpha)} (t, g)
\left( \ln Q ( \pm \ft{i}{2} - g t ) \right)^\prime
+
\theta^{(\alpha)} (t, g)
\left(
\ln
\frac{
Q ( + \ft{i}{2} - g t )
}{
Q ( - \ft{i}{2} - g t )
}
\right)^\prime
\right\}
\, ,
\ee
with the weight functions being decomposed in terms of the Chebyshev polynomials $U_k(t)$,
\ba
&&
w^{(1)} (t, g) = - g^2
\, , \qquad
\vartheta^{(1)} (t, g) =
\ft{i}{2} g^2 \sum_{s = 3}^\infty (- 1)^s C_{2s} (g) U_{s - 2} (t)
\, , \\
&& w^{(2)} (t, g) = \ft12 g^3 t
\, , \qquad
\vartheta^{(2)} (t, g)
=
i g^3 \sum_{s = 2}^\infty (- 1)^s C_{3s} (g) U_{s - 2} (t)
\, . \nonumber
\ea

Having the functional form of the transfer matrix fixed, one can immediately solve the Baxter
equation order by order of perturbation theory by determining the roots of the $Q$-polynomial
in terms of the charges $\mathfrak{Q}_k^{[\ell]}$ and $\mathfrak{R}_k^{[\ell]}$ and then
finding the latter. However, one can make a step further and completely constrain the
compensatory charges at $\ell$-th order in the 't Hooft coupling in terms of Baxter functions
$Q_{\ell^\prime} (u)$ at orders $\ell^\prime < \ell$. We present below solutions for
up to twist four and six loops. As a demonstration of the efficiency of the framework,
we compute a few anomalous dimensions including wrapping contributions from the L\"uscher
formula \re{WrappingGamma}.

{\bf 3. Twist two.} Let us discuss the twist two and start by noticing that in the first
three orders in the 't Hooft coupling $g^2$, the charges $\mathfrak{Q}_{k > 2}^{[\ell < 4]} = 0$
all vanish, while $\mathfrak{R}_k$ emerge already from four loops according to \re{NonvanishR},
which is also the order when one has to account for wrapping effects. A straightforward
calculation yields the following results for nonvanishing compensatory charges expressed in
terms of the $(\ell + 1)$-loop contribution $\gamma_\ell^+$ to the anomalous dimensions
\re{AsyAD} and derivatives of the Baxter function summarized in the Appendix.
\begin{itemize}
\item Four loops:
\ba
\mathfrak{R}_1^{[3]} \!\!\!&=&\!\!\!
- \ft12 \Re{\rm e} \left[ \gamma^+_0 ( (\gamma^+_0)^2 + \alpha^+ ) \right]
\, , \\
\mathfrak{R}_2^{[3]} \!\!\!&=&\!\!\!
0
\, , \nonumber\\
\mathfrak{R}_3^{[3]} \!\!\!&=&\!\!\!
\ft1{8} \Re{\rm e} \left[ \gamma^+_0 \right]
\, . \nonumber
\ea
\item Five loops:
\ba
\mathfrak{R}_1^{[4]} =
\!\!\!&-&\!\!\!
\Re{\rm e}
\big[
\ft12 \gamma^+_1 \left[ 3 (\gamma^+_0)^2 + \alpha^+ \right]
\\
&+&\!\!\!
\ft{1}{8}
\gamma^+_0
\left[
4 (\gamma^+_0)^4 + 4 (\gamma^+_0)^2 \alpha^+ - (\alpha^+)^2 + 16 \chi^+ - 4 \omega^+
\right]
\big]
-
\ft12 \zeta_3 \Re{\rm e} \left[ (\gamma^+_0)^2 \right]
\, , \nonumber\\
\mathfrak{R}_2^{[4]} \!\!\!&=&\!\!\!
0
\, , \nonumber\\
\mathfrak{R}_3^{[4]} \!\!\!&=&\!\!\!
\ft{1}{8} \Re{\rm e} \left[ \gamma^+_1 \right]
\, , \nonumber
\ea
and the $\mathfrak{Q}^{[4]}_{k > 2}$ coefficients:
\ba
\mathfrak{Q}_{3}^{[4]}
\!\!\!&=&\!\!\! 0
\, , \\
\mathfrak{Q}_{4}^{[4]}
\!\!\!&=&\!\!\!
\Re{\rm e}
\left[
\ft3{32} (\alpha^+)^2
-
\ft1{24}
(\gamma^+_0)^2
\left[
5 (\gamma^+_0)^2 + 3 \alpha^+ + 2 \beta^+
\right]
-
\ft1{24} \gamma^+_0 [2 \gamma_1^+ - \varepsilon^+] - \ft1{8} \omega^+
\right]
\, , \nonumber\\
\mathfrak{Q}_{5}^{[4]}
\!\!\!&=&\!\!\! 0
\, , \\
\mathfrak{Q}_{6}^{[4]}
\!\!\!&=&\!\!\!
\ft1{16} \Re{\rm e} \left[ (\gamma^+_0)^2 \right]
\, . \nonumber
\ea
\item Six loops:
\ba
\mathfrak{R}_5^{[5]}
\!\!\!&=&\!\!\!
\ft{1}{48}
\Re{\rm e}
\left[
\gamma_1^+ + \gamma_0^+ \beta^+ - 2 (\gamma_0^+)^3 - \ft12 \varepsilon^+
\right]
\, , \\
\mathfrak{R}_4^{[5]}
\!\!\!&=&\!\!\!
0 \, , \nonumber\\
\mathfrak{R}_3^{[5]}
\!\!\!&=&\!\!\!
\ft{1}{24}
\Re{\rm e}
\big[
6 \gamma_2^+
- 2 \gamma_1^+ [(\gamma_0^+)^2 + \alpha^+]
\nonumber\\
&+&\!\!\! (\gamma_0^+)^2
\left[ 4 (\gamma_0^+)^3 - 2 (\gamma_0^+) (\beta^+ - 2 \alpha^+) + \varepsilon^+ \right]
- \alpha^+ (2 \gamma_0^+  \beta^+ +  \varepsilon^+)
\big]
\, , \nonumber\\
\mathfrak{R}_2^{[5]}
\!\!\!&=&\!\!\!
0 \, , \nonumber\\
\mathfrak{R}_1^{[5]}
\!\!\!&=&\!\!\!
-
\Re{\rm e}
\big[
\ft{1}{2} \gamma_2^+ [3 (\gamma_0^+)^2 + \alpha^+]
+ \ft{29}{18} (\gamma_1^+)^2 \gamma_0^+
\nonumber\\
&+&\!\!\!
\ft{1}{36} \gamma_1^+
\left[ 104 (\gamma_0^+)^4 + 66 (\gamma_0^+)^2 \alpha^+ + 8 (\gamma_0^+)^2 \beta^+
- 4 \gamma_0^+ \varepsilon^+ - 3 (3 (\alpha^+)^2 - 24 \chi^+ + 4 \omega^+ ) \right]
\nonumber\\
&+&\!\!\!
\ft{7}{9} (\gamma_0^+)^7
+
\ft{1}{18} (\gamma_0^+)^5 \left[ 15 \alpha^+ + 7 \beta^+ \right]
\nonumber\\
&+&\!\!\!
\ft{1}{36} (\gamma_0^+)^3 \left[ 72 \chi^+ - 9 (\alpha^+)^2 +12 \alpha^+ \beta^+ + 4 (\beta^+)^2 + 6 \omega^+ \right]
\nonumber\\
&-&\!\!\!
\ft{1}{36} (\gamma_0^+)^2 \left[ 576 \xi^+ + (7 (\gamma_0^+)^2 + 6 \alpha^+ + 4 \beta^+ ) \varepsilon^+ \right]
\nonumber\\
&-&\!\!\!
\ft{1}{72} \gamma_0^+
\left[
9 \alpha^+ (8 \chi^+ - 6 \omega^+ + (\alpha^+)^2 + \alpha^+ \beta^+)
+ 2 ( 72 \eta^+ + 288 \kappa^+ - 24 \beta^+ (\omega^+ - 3 \chi^+) )
\right]
\nonumber\\
&-&\!\!\!
\ft{1}{144} (12 \omega^+ - 9 (\alpha^+)^2 - 4 \gamma_0^+ \varepsilon^+) \varepsilon^+
\big]
\nonumber\\
&-&\!\!\! \zeta_3 \, \Re{\rm e}
\left[
\gamma_0^+ \left( \gamma_1^+ + \ft12 \gamma_0^+ \alpha^+ + \ft12 (\gamma_0^+)^3 \right)
\right]
+ \ft{9}{8} \zeta_5 \, \Re{\rm e} \left[ (\gamma_0^+ )^2 \right]
\, , \nonumber
\ea
and the $\mathfrak{Q}^{[5]}_{k > 2}$:
\ba
\mathfrak{Q}_{3}^{[5]}
\!\!\!&=&\!\!\! 0
\, , \\
\mathfrak{Q}_{4}^{[5]}
\!\!\!&=&\!\!\!
- \Re{\rm e}
\big[
\ft{1}{12} (\gamma_1^+)^2
-
\ft{1}{24} \gamma_1^+ (\varepsilon^+ + 2 \gamma_0^+ \alpha^+ - 12 (\gamma_0^+)^3)
-
\ft{1}{12} (\gamma_0^+)^6
-
\ft{1}{6} (\gamma_0^+)^4 (2 \alpha^+ + \beta^+)
\nonumber\\
&+&\!\!\!
\ft{1}{24} (\gamma_0^+)^3 \varepsilon^+
-
\ft{1}{24} (\gamma_0^+)^2
\left( 3 (\alpha^+)^2 + 8 \alpha^+ \beta^+ + 2 (\beta^+)^2 + 6 \omega^+ \right)
\nonumber\\
&+&\!\!\!
\ft{1}{48}
\gamma_0^+
\left(
5 \alpha^+ \varepsilon^+ + 2 \beta^+ \varepsilon^+ + 192 \xi^+ + 48 \phi^+
\right)
\nonumber\\
&+&\!\!\!
\ft{1}{16}
\left(
3 (\alpha^+)^3 + 32 \kappa^+ - 2 \beta^+ \omega^+ - 2 \alpha^+ (6 \chi^+ + \omega^+)
\right)
\big]
\nonumber\\
&-&\!\!\!
\ft18 \zeta_3 \, \Re{\rm e} \left[\gamma_0^+ ( (\gamma_0^+)^2 - \alpha^+ ) \right]
+ \ft{1}{32} \zeta_5 \Re{\rm e} \left[ \gamma_0^+ \right]
\, , \nonumber\\
\mathfrak{Q}_{5}^{[5]}
\!\!\!&=&\!\!\! 0
\, , \nonumber\\
\mathfrak{Q}_{6}^{[5]}
\!\!\!&=&\!\!\!
\ft1{8} \Re{\rm e} \left[ \gamma_1^+ \gamma^+_0 \right]
\, . \nonumber
\ea
\end{itemize}
We simplified these expressions making use of the fact that only zero quasimomentum states
are physical such that taking the real part of these expressions becomes redundant as the
Baxter polynomial is an even function with resect to the spectral parameter $u$. However
we will keep the real part operation here for conformity with other values of $L$ presented
below.

The revised Baxter equation can be easily diagonalized for specific values of the conformal
spin. For example, for $N = 10$ one finds at four-, five- and six-loops
\ba
\gamma_3
\!\!\!&=&\!\!\!
- \ft{1605938976900071848052623}{64536258792112128000000}
- \ft{83423671309}{45722880000} \zeta_3
\\
\!\!\!&-&\!\!\!
\ft{54479161}{101606400}
\left(
40 \zeta_5 - \ft{5194387}{198450} \zeta_3 - \ft{70616856388243}{7201353600000}
\right)
\, , \nonumber\\
\gamma_4^{(\rm asy)}
\!\!\!&=&\!\!\!
\ft{18166776040401946073522569013807}{292736469881020612608000000000}
+
\ft{413968341940646383}{58071715430400000} \zeta_3
+
\ft{83423671309}{18289152000} \zeta_5
\, , \\
\gamma_5^{(\rm asy)}
\!\!\!&=&\!\!\!
-
\ft{103472529189518339662036825500201747281}{619664559444144432768614400000000000}
\\
&-&\!\!\!
\ft{1820094012031680420099427}{76828879514419200000000} \zeta_3
-
\ft{21141203559973933733}{1161434308608000000} \zeta_5
-
\ft{(83423671309}{6967296000} \zeta_7
\, , \nonumber
\ea
respectively. Notice that at four loops, the second line incorporates the contribution
from the wrapping effects in agreement with earlier findings of Ref.\ \cite{BajJanLuk08}.

{\bf Twist three.} The analysis of the twist-three sector goes along the same lines as
outlined above. Compared to the twist two case, the compensatory charges $\mathfrak{Q}_{k > L}$
emerge starting from two loops already and are related to the lower-order spin-chain charges
\be
\mathfrak{Q}_4^{[\ell]} = \ft14 \mathfrak{Q}_2^{[\ell - 1]}
\, , \quad
\ell = 1,2,3
\, .
\ee
Nontrivial effects appear from five loops and on. First, in $\mathfrak{Q}_3^{[4]}$ there is
an additive correction to the charge $ \mathfrak{Q}_2^{[3]}$,
\begin{itemize}
\item Five loops:
\be
\mathfrak{Q}_4^{[4]} = \ft14 \mathfrak{Q}_2^{[3]} + \mathfrak{S}_4^{[4]}
\, ,
\ee
with the latter being
\ba
\mathfrak{S}_4^{[4]}
\!\!\!&=&\!\!\!
\ft{1}{8} \zeta_3 \, \Re{\rm e} [ \gamma_0^+ ]
+
\Re{\rm e}
\big[
\ft{1}{24} \left( \gamma_0^+ + \gamma_0^- \right)
\left(
4 \gamma_1^+ - ( \gamma_0^+ + \gamma_0^-)^3 + \varepsilon^+
\right)
+
\ft{1}{12} \gamma_0^+ \left( \gamma_0^+ \beta^+ + 3 \gamma_1^+ \right)
\nonumber\\[1mm]
&+&\!\!\!
\ft{1}{48} \left( \beta^+ + \beta^- \right)
\left(
\gamma_0^+ \gamma_0^- + 2 ( \gamma_0^+ )^2 + 3 \alpha^+
\right)
+
\ft{1}{32} \left( \alpha^+ \alpha^- - 2 (\alpha^+)^2 \right)
+
\ft{1}{2} \chi^+
-
\ft{1}{8} \omega^+
\big]
\, . \nonumber\\
\ea
The rest of nonvanishing compensatory coefficients read
\ba
\mathfrak{R}_1^{[4]}
\!\!\!&=&\!\!\!
- \ft{1}{24}
\Im{\rm m}
\left[
\left( \gamma_0^+ - \gamma_0^- \right)
\left( 2 \gamma_1^+ + ( \gamma_0^+ + \gamma_0^- )^3 + 2 \gamma_0^+ \beta^+ - \varepsilon^+
\right)
\right]
\, , \\
\mathfrak{R}_2^{[4]}
\!\!\!&=&\!\!\!
0 \, , \nonumber\\
\mathfrak{R}_3^{[4]}
\!\!\!&=&\!\!\!
- \ft{1}{16} \Im{\rm m} \left[  (\gamma_0^+)^2 \right]
\, , \nonumber\\
\mathfrak{R}_4^{[4]}
\!\!\!&=&\!\!\!
\ft{1}{32} \Re{\rm e} \left[ \gamma_0^+ \right]
\, . \nonumber
\ea
These formulas were significantly simplified making use of the following identity
\be
2 \gamma_0^+ \gamma_0^- - 2 (\gamma_0^+ + \gamma_0^-)^2 - \alpha^+ - \alpha^- = 0
\, .
\ee
The latter does not reduce, however, the complexity of the six-loop expressions such that
$\mathfrak{Q}_k^{[5]}$ and $\mathfrak{R}_k^{[5]}$ are rather lengthy to be presented
here.
\end{itemize}

These findings were used to find selected anomalous dimensions. For instance, for $N = 6$
there are two eigenvalues with zero quasimomentum. One of them possesses a nontrivial value
of the charge $\mathfrak{Q}_3^{[0]}$ and thus signals the appearance of an excited state.
Their anomalous dimensions are
\ba
\gamma_3 [0]
\!\!\!&=&\!\!\!
- \ft{83243489}{8957952} - \ft{2761}{2592} \zeta_3
\, , \\ [1mm]
\gamma_3 [ \pm 2 \sqrt{723}]
\!\!\!&=&\!\!\!
- \ft{20718946661183}{884736000000} - \ft{2346601}{1536000} \zeta_3
\, ,
\ea
at four loops and
\ba
\gamma_4 [0]
\!\!\!&=&\!\!\!
\ft{8407309337}{429981696} + \ft{1444117}{373248} \zeta_3
+ \ft{13805}{5184} \zeta_5
\nonumber\\
&-&\!\!\!
\ft{121}{576}
\left(
35 \zeta_7 - \ft{490}{9} \zeta_5 + \ft{2401}{162} \zeta_3 + \ft{613793}{139968}
\right)
\, , \\ [1mm]
\gamma_4 [\pm 2 \sqrt{723}]
\!\!\!&=&\!\!\!
\ft{626236747334313089}{10616832000000000}
+
\ft{21046079153}{3686400000} \zeta_3
+
\ft{2346601}{614400} \zeta_5
\nonumber\\
&-&\!\!\!
\ft{51529}{102400}
\left(
35 \zeta_7 + \ft{4151}{120} \zeta_5 - \ft{353563}{8000} \zeta_3
- \ft{136100137}{7680000}
\right)
\, ,
\ea
five loops, respectively, with numbers in the square brackets showing the value of the
conserved charge $\mathfrak{Q}_3^{[0]}$. The second lines in the last two equations come
from the L\"uscher correction \re{WrappingGamma}, while the first lines stem from the
asymptotic Baxter equation. The eigenvalue with zero $\mathfrak{Q}_3^{[0]}$ coincides
with the result of Ref.\ \cite{BecForLukZie09}. Finally, at six loops we get the asymptotic
anomalous dimensions for the same states,
\ba
\gamma_5^{(\rm asy)} [0]
\!\!\!&=&\!\!\!
- \ft{2720281112987}{61917364224}
- \ft{(317437583}{26873856} \zeta_3
- \ft{793787}{82944} \zeta_5
- \ft{(96635}{13824} \zeta_7
\, , \\
\gamma_5^{(\rm asy)} [\pm 2 \sqrt{723}]
\!\!\!&=&\!\!\!
-
\ft{2742831213855911614841}{16986931200000000000}
-
\ft{42825125493751}{2359296000000} \zeta_3
-
\ft{430388898599}{29491200000} \zeta_5
-
\ft{16426207}{1638400} \zeta_7
\, .
\ea

{\bf Twist four.} Lastly, we turn to the $L=4$ operators. For the lowest three orders of
perturbation theory, we find the following identities
\be
\mathfrak{Q}_5^{[\ell]} = \ft14 \mathfrak{Q}_3^{[\ell - 1]}
\, , \quad
\mathfrak{Q}_6^{[\ell]} = \ft1{16} \mathfrak{Q}_2^{[\ell - 2]}
\, , \qquad
\ell = 1,2,3
\, ,
\ee
which are rather obvious generalization of the ones for the $L=3$ case. While at higher loops
they acquire additional corrections\footnote{Obviously, $\mathfrak{Q}^{[\ell < 0]} = 0$.}
\be
\mathfrak{Q}_5^{[\ell]}
= \ft14 \mathfrak{Q}_3^{[\ell - 1]}
+ \mathfrak{S}_5^{[\ell]}
\, , \quad
\mathfrak{Q}_6^{[\ell]} = \ft1{16} \mathfrak{Q}_2^{[\ell - 2]}
+ \mathfrak{S}_6^{[\ell]}
\, , \qquad
\ell > 3
\, ,
\ee
which will be given below order-by-order in coupling.
\begin{itemize}
\item Five loops:
\ba
\mathfrak{S}_5^{[4]}
\!\!\!&=&\!\!\!
0
\, , \\
\mathfrak{S}_6^{[4]}
\!\!\!&=&\!\!\!
\ft{1}{32}
\Re{\rm e}
\left[ 2 (\gamma^+_0)^2 + \gamma^+_0 \gamma^-_0 + \alpha^+ \right]
\, , \nonumber
\ea
where the following identity was employed
\be
2 \gamma_1^+ + 2 \gamma_1^- + 2 \gamma_0^+ \beta^+ + 2 \gamma_0^- \beta^-
+ \varepsilon^+ + \varepsilon^-
=
(\gamma_0^+ + \gamma_0^-)
\left[
4 (\gamma_0^+ + \gamma_0^-)^2 - 6 \gamma_0^+ \gamma_0^- + 3 \alpha^+ + 3 \alpha^-
\right]
\, .
\ee
\item Six loops:
\ba
\mathfrak{S}_5^{[5]}
\!\!\!&=&\!\!\!
- \ft{1}{16} \zeta_3 \Im{\rm m} \left[ ( \gamma_0^+ )^2 \right]
-
\ft{1}{24}
\Im{\rm m}
\Big[
\gamma_1^+
\left(
9 (\gamma_0^+)^2 + 2 (\gamma_0^-)^2 + 5 \gamma_0^+ \gamma_0^-
+ \alpha^- + 3 \alpha^+ + \beta^+
\right)
\\
&&\qquad\qquad\qquad \,
+
(\gamma_0^+)^3
\left( 3 ( \gamma_0^+ )^2 + 2 \gamma_0^+ \gamma_0^- + 2 (\gamma_0^-)^2 \right)
\nonumber\\[1mm]
&+&\!\!\!
\ft{1}{2}
(\gamma_0^+)^2
\left(
\gamma_0^-
\left( 3 \alpha^+ + 3 \alpha^- - 2 \beta^+ - 2 \beta^- \right)
+
2
\gamma_0^+
\left( 3 \alpha^+ + \alpha^- + \beta^+ \right)
+
\varepsilon^-
\right)
\nonumber\\[1mm]
&+&\!\!\!
\ft{1}{2}
\gamma_0^+
\left(
\ft32 (\alpha^+ - \alpha^-)^2 + \ft32 \alpha^+ \alpha^- - 24 \chi^+ - 12 \chi^-
+ ( \alpha^- - 3 \alpha^+ - 2 \beta^+ ) \beta^+ + 3 \omega^+ + 3 \omega^-
\right)
\nonumber\\
&&\qquad\qquad\qquad \,
+
\ft{1}{4}
\left( 2 \gamma_0^+ \gamma_0^- + \alpha^- - 2 \beta^+ \right) \varepsilon^+
+ 48 \xi^+ - 12 \phi^+
\Big]
\, , \nonumber\\
\mathfrak{S}_6^{[5]}
\!\!\!&=&\!\!\!
\ft{1}{32} \zeta_3 \, \Re{\rm m} \left[ \gamma_0^+ \right]
\nonumber\\
&+&\!\!\!
\ft{1}{96}
\Re{\rm m}
\Big[
\gamma^+_0 ( 14 \gamma_1^+ + 6 \gamma_1^- - \varepsilon^+ )
+
(\gamma^+_0 )^2 \left( (\gamma^+_0)^2 + \beta^+ \right)
- 3 (\alpha^+)^2 + 12 \chi^+
\Big]
\, , \nonumber
\ea
and
\ba
\mathfrak{R}_1^{[5]}
\!\!\!&=&\!\!\!
\ft{1}{24}
\Re{\rm e}
\Big[
\gamma_1^+
\left(
(3 \gamma_0^+ + 2 \gamma_0^-)(\gamma_0^- - \gamma_0^+)
+ \alpha^- - \alpha^+ + \beta^- - \beta^+
\right)
\\
&&\qquad\qquad\qquad \,
+
(\gamma_0^+)^3
\left( 3 ( \gamma_0^+ )^2 + 4 \gamma_0^+ \gamma_0^- + 2 (\gamma_0^-)^2
+ 5 \alpha^+ + \alpha^- - \beta^+ - 2 \beta^-
\right)
\nonumber\\[1mm]
&+&\!\!\!
\ft{1}{2}
(\gamma_0^+)^2 \gamma_0^-
\left(
9 \alpha^+ - 3 \alpha^-
-
10 \beta^- + 8 \beta^+
\right)
\nonumber\\[1mm]
&+&\!\!\!
\ft{1}{2}
\gamma_0^+
\big(
\beta^+ ( \alpha^+ + 5 \alpha^- - 2 \beta^+)
-
\beta^- ( 3 \alpha^+ + 3 \alpha^- - 2 \beta^+)
-
3 \omega^+ - 3 \omega^-
\nonumber\\
&&\qquad\qquad\qquad \,
+
\ft92 (\alpha^+)^2
+
\ft32 \alpha^+ \alpha^-
-
\ft32 (\alpha^-)^2
\big)
\nonumber\\
&-&\!\!\!
\ft{1}{2}
\left(
\gamma_0^+ \gamma_0^- -3 (\gamma_0^+)^2 + 2 (\gamma_0^-)^2 - \beta^+ - \beta^- - \alpha^+ + \alpha^-
\right) \varepsilon^+
\Big]
\, , \nonumber\\
\mathfrak{R}_2^{[5]}
\!\!\!&=&\!\!\!
0
\, , \nonumber\\
\mathfrak{R}_3^{[5]}
\!\!\!&=&\!\!\!
- \ft{1}{64}
\Re{\rm e}
\Big[
2 (\gamma_0^+)^2 \left( 2 \gamma_0^+ + \gamma_0^- \right)
- \gamma_0^+ \left( \alpha^- - 3 \alpha^+ - 2 \beta^+ + 2 \beta^- \right)
\Big]
\, , \nonumber\\
\mathfrak{R}_4^{[5]}
\!\!\!&=&\!\!\!
- \ft{1}{64}
\Re{\rm e} \left[ 2 (\gamma_0^+)^2 + \alpha^+ + \beta^+ \right]
\, , \nonumber\\
\mathfrak{R}_5^{[5]}
\!\!\!&=&\!\!\!
\ft{1}{128}
\Re{\rm e} \left[ \gamma_0^+ \right]
\, . \nonumber
\ea
\end{itemize}
As a demonstration, we computed the eigenvalues of the $L=4$, $N=2$ Wilson operator at
five loops\footnote{The rational part of the five-loop anomalous dimension is in agreement
with the asymptotic Bethe Ansatz prediction for BMN $L=6,\ N=2$ eigenstates which enter
into the same supermultiplet with the considered Wilson operator \cite{BeiDipSta04}.}
\ba
\gamma_4 [0, \pm \sqrt{5}]
\!\!\!&=&\!\!\!
\ft{263125 \pm 109793 \sqrt{5}}{12800}
+ \ft{1}{64} (69 \pm 19 \sqrt{5}) \zeta_3
+ \ft{5}{32} (5 \pm \sqrt{5}) \zeta_5
\, .
\ea
and at six loops
\ba
\gamma_5 [0, \!\!\!&\pm&\!\!\! \sqrt{5}]
=
- \ft{2898675 \pm 1249357 \sqrt{5}}{51200}
- \ft{(4535 \pm 1541 \sqrt{5})}{1280} \zeta_3
- \ft{15}{128} (24 \pm 7 \sqrt{5}) \zeta_5 - \ft{105}{256} (5 \pm \sqrt{5}) \zeta_7
\nonumber\\
&-&\!\!\!
\ft{5}{128} (3 \pm \sqrt{5})
\left(
\ft{63}{2} \zeta_9
- \ft{7}{4} ( 5 \mp 7 \sqrt{5} ) \zeta_7
+ \ft{5}{2} ( 1 \mp 2 \sqrt{5} ) \zeta_5
- \ft{5}{2} ( 5 \pm 2 \sqrt{5} ) \zeta_3
- \ft{25}{32} ( 7 \pm 3 \sqrt{5} )
\right)
\, , \nonumber\\
\ea
with the wrapping correction displayed in the second line. Again the one-loop eigenvalues
of the conserved charges $\mathfrak{Q}_3^{[0]}$ and $\mathfrak{Q}_4^{[0]}$ are shown in the
square brackets.

{\bf 6. Outlook.} The formalism we have presented here allows one to find analytical form
of anomalous dimensions at high orders of perturbation theory for given conformal spins
$(N + \ft12 L)$ which can be arbitrarily large. We have limited ourselves to just a few
specific eigenvalues merely due to space limitation. There are several avenues where the
modified Baxter equation, we have presently suggested, can be employed in an efficient
fashion. First, it can be used for analytical calculation of the anomalous dimensions of
low-twist operators along the lines of Ref.\ \cite{KotRejZie08}. Next, its asymptotic
expansion for large spins $N$, as developed in Ref.\ \cite{BelKorPas08}, can immediately
be used to unravel the structure of preasymptotic terms, including nontrivial effects
stemming from wrapping. These and other questions will be addressed elsewhere.

\vspace{5mm}

\noindent This work was supported by the U.S. National Science Foundation under grant no.\
PHY-0456520. We would like to thank Stefan Zieme for renewing the author's interest in
questions discussed here, Romuald Janik for usefull correspondence and Yuri Dokshitzer,
Gavin Salam and Gregory Korchemsky for hospitality at Jussieu and Saclay, respectively, in
December of 2008 when this work was initiated.

\vspace{5mm}

\appendix

{\bf Appendix.} In the main body of the paper, we have used the following notations
in order to write down all results in a concise fashion:
\baa
\alpha^+
\!\!\!&=&\!\!\!
\ft14 \frac{Q_0^{\prime\prime} (\ft{i}{2})}{Q_0 (\ft{i}{2})}
\, , \qquad\qquad\qquad
\beta^+
=
\frac{Q_1 (\ft{i}{2})}{Q_0 (\ft{i}{2})}
\, , \\
\varepsilon^+
\!\!\!&=&\!\!\!
i \frac{Q_1^\prime (\ft{i}{2})}{Q_0 (\ft{i}{2})}
\, , \qquad\qquad\qquad \
\omega^+
=
\ft{1}{192} \frac{Q_0^{(4)} (\ft{i}{2})}{Q_0 (\ft{i}{2})}
\, , \\
\xi^+
\!\!\!&=&\!\!\!
\ft{i}{61440}
\frac{Q_0^{(5)} (\ft{i}{2})}{Q_0 (\ft{i}{2})}
\, , \qquad\qquad
\kappa^+
=
\ft{1}{3072}
\frac{Q_1^{(4)} (\ft{i}{2})}{Q_0 (\ft{i}{2})}
+
\ft{1}{61440}
\frac{Q_0^{(6)} (\ft{i}{2})}{Q_0 (\ft{i}{2})}
\, , \\
\chi^+
\!\!\!&=&\!\!\!
\ft{1}{16} \frac{Q_1^{\prime\prime} (\ft{i}{2})}{Q_0 (\ft{i}{2})}
-
\ft{1}{16}
\frac{Q_1 (\ft{i}{2}) Q_0^{\prime\prime} (\ft{i}{2})}{Q^2_0 (\ft{i}{2})}
+
\ft{1}{192}
\frac{Q_0^{(4)} (\ft{i}{2})}{Q_0 (\ft{i}{2})}
\, , \\
\phi^+
\!\!\!&=&\!\!\!
\ft{i}{192} \frac{Q_1^{(3)} (\ft{i}{2})}{Q_0 (\ft{i}{2})}
+
\ft{i}{3072} \frac{Q_0^{(5)} (\ft{i}{2})}{Q_0 (\ft{i}{2})}
\, , \\
\eta^+
\!\!\!&=&\!\!\!
- \ft{1}{16} \frac{Q_2^{\prime\prime} (\ft{i}{2})}{Q_0 (\ft{i}{2})}
+ \ft{1}{16} \frac{Q_2 (\ft{i}{2}) Q_0^{\prime\prime} (\ft{i}{2})}{Q^2_0 (\ft{i}{2})}
- \ft{1}{192} \frac{Q_1^{(4)} (\ft{i}{2})}{Q_0 (\ft{i}{2})}
- \ft{1}{6144} \frac{Q_0^{(6)} (\ft{i}{2})}{Q_0 (\ft{i}{2})}
\, .
\eaa

%%%%%%%%%%%%%%%%%%%%%%%%%%%%%%%%%%%%%%%%%%%%%%%%%%%%%%%%%%%%%%%%%%%%%

%%%%%%%%%%%%%%%%%%%%%%%%%%%%%%%%%%%%%%%%%%%%%%%%%%%%%%%%%%%%%%%%%%%%%

\end{document}